\documentclass[12pt,a4paper]{article}
\usepackage[english]{babel}
\usepackage{subcaption} 
\usepackage{wrapfig}
\usepackage{graphicx}
\usepackage{epstopdf}
\usepackage{multicol}
\usepackage{tabularx}
\usepackage{amssymb,amsmath}
\usepackage{mathrsfs}
\usepackage{xfrac}
\usepackage{textcomp}
\usepackage{bm}
\usepackage{cite}
\usepackage{color}
\usepackage{ulem}
\usepackage{setspace} 
\usepackage{float}
\newcommand{\onehf}{\mbox{\sfrac{1}{2}}} 

\definecolor{dblue}{rgb}{0,0,0.8}

\definecolor{dred}{rgb}{0.7,0,0}

\definecolor{dgreen}{rgb}{0.,0.6,0}

\makeatletter 
\def\lambdabar{\protect\@lambdabar}
\def\@lambdabar{%
\relax
\bgroup
\def\@tempa{\hbox{\raise.73\ht0
\hbox to0pt{\kern.25\wd0\vrule width.5\wd0
height.1pt depth.1pt\hss}\box0}}%
\mathchoice{\setbox0\hbox{$\displaystyle\lambda$}\@tempa}%
{\setbox0\hbox{$\textstyle\lambda$}\@tempa}%
{\setbox0\hbox{$\scriptstyle\lambda$}\@tempa}%
{\setbox0\hbox{$\scriptscriptstyle\lambda$}\@tempa}%
\egroup
}
\makeatother

\begin{document}

\begin{center}
{\Large {\bf On classical and quantum effects at scattering of fast charged particles in ultrathin crystal}}

\vskip 0.5cm

{\bf S.N.~Shul'ga$^{a}$, N.F.~Shul'ga$^{a,b}$, S.~Barsuk$^{c}$, I.~Chaikovska$^{c}$, R.~Chehab$^{c}$}

\vskip 0.3cm

$^{a}${\it NSC ``Kharkov Institute of Physics and Technology", Kharkiv 61108, Ukraine}

$^{b}${\it Karazin Kharkiv National University, Kharkiv 61022, Ukraine}

$^{c}${\it  LAL (IN2P3-CNRS and Paris Sud University), 91898 Orsay C\'edex, France}

\end{center}

\vskip 0.4cm

\begin{abstract}
\noindent Classical and quantum properties of scattering of charged particles in ultrathin crystals are considered. A comparison is made of these two ways of study of scattering process. In the classical consideration we remark the appearance of sharp maxima that is referred to the manifestation of the rainbow scattering phenomenon and in quantum case we show the sharp maxima that arise from the interference of single electrons on numerous crystal planes, that can be expressed in the terms of reciprocal lattice vectors. We show that for some parameters quantum predictions substantially differ from the classical ones. Estimated is the influence of the beam divergence on the possibility of experimental observation of the studied effects.
\end{abstract}

\noindent

\vskip 0.4cm


\vskip 0.4cm


\vspace{-2ex}
\section{Introduction}
\vspace{-2ex}

The motion of a fast charged particle near direction of one of its planes or axes can be considered as a motion in the field of continuous planes or strings. These are the cases of particle channelling or over-barrier motion. A number of theoretical and experimental studies have been made devoted to these phenomena (see, e.g., \cite{Lindh1965,Gemmell1974,Kumakh1986,BazZhev1987,AkhiezShulga_HighEn1996,RArtDh98} and references therein). In order to describe the effects of interaction of a charged particle with medium we must get first of all the characteristics of its motion.

Interesting phenomena may happen just at the beginning of such motion, in the transitional area before the particle has completed several oscillations inherent in channelling or above-barrier motion in this case. Such a regime is realized in crystals thin enough, with the thicknesses that vary from less than tenths of micron (hundreds of \AA) for MeV particles to several tens of microns for hundreds of GeV particles (the characteristic dimension of such an area depends on the particle energy as a square-root function). In our study we will call these ultrathin crystals. In this work we will mostly consider few-MeV charged particles, so our range of crystal thicknesses spreads from about hundreds up to thousands of {\AA}ngstr\"oms. In the last years the technologies were developed to produce such crystals, and these crystals have been used for channelling experiments \cite{Guidi,Motapoth2,MotapothNesk_PRB2012,Motapoth1,Motapoth3}. In the paper we will propose the experimental study of angular distributions of electrons scattered by an unltrathin crystal. 

The problem of obtaining characteristics of the motion of a charged particle in these conditions may be resolved both by means of quantum and classical theories, at that higher is the particle energy, more the quantum and classical solutions match one other.

In this work we will stress on the energies low enough so as quantum effects become essential in the particle motion but still high enough so that the crystal thickness required for observing these effects was reachable. For electrons we can propose the study at energies 4\,Mev for mostly quantum motion and 50\,MeV where some comparison of quantum results with the classical ones begins to be reasonable. The conditions, necessary for the study of phenomena considered in this paper, can be met with use of modern technologies of creating ultrathin crystals and the experiments can be realized on the base of accelerators PhIL and ThomX in the laboratory LAL in Orsay, France.

In this paper we consider the case of planar scattering only as the one which reveals the essence of the nature of the processes, two-dimensional case of axial scattering being mostly only a generalization of it (although, some phenomena as, for example, dynamical chaos, will be only possible at axial scattering). Our observations require low beam divergence, so one should aspire to get it low for experiments. It is possible to improve the divergence by squeezing the beam in the direction perpendicular to crystal planes using magnetic field: simultaneously the beam will stretch out along the planes but this would create no problem in our one-dimensional study.

Here we will consider the case of parallel incidence of a particle relatively a crystal plane. The incidence under a small angle relatively plane reveals other interesting effects and is a subject of a separate study.

\vspace{-2ex}
\section{Classical scattering}
\vspace{-2ex}

We can consider the interaction of a fast particle with matter both within the classical and the quantum theory. The classical theory of scattering is based upon the definition of the particle trajectory in external field. At motion of a particle along crystal planes its trajectory in transversal direciton is defined as a solution of the differential equation of motion \cite{Lindh1965,Gemmell1974,Kumakh1986,BazZhev1987,AkhiezShulga_HighEn1996},
\begin{equation}\label{eq_clasrho}
\ddot{x}  =  - \frac{\,\,{c^2}}
{E_\parallel}\frac{\partial}{\partial x} U\left( x\right),
\end{equation}
where $x$ is the coordinate of the particle in the plane of transverse motion, ${E_\parallel} = c\sqrt{p_\parallel^2 +{m^2}{c^2}}$ and $U(x)$ is the potential energy of a particle in the continuous potential of crystal planes. In this article we will neglect such phenomena as multiple scattering or radiation energy losses of a particle by assuming them to be small enough, that is the consequence of a small crystal thickness, therefore small particle path in the field of atomic forces. 

The continuous potential of a crystal plane is obtained as the average of the fields of atoms along it with taking into account of random deviations of atom positions relatively their places in the lattice caused by heat oscillations \cite{AkhiezShulga_HighEn1996}. As a model of a solitary atom potential we took the Moli\`ere potential that is widely used to describe atomic electric forces. In order to obtain the potential of entire crystal along the chosen direction we must summarize all non-negligible contributions of the neighbouring planes.

\begin{wrapfigure}[21]{l}{0.5\linewidth} 
\vspace{-3ex}
\includegraphics[width=\linewidth]{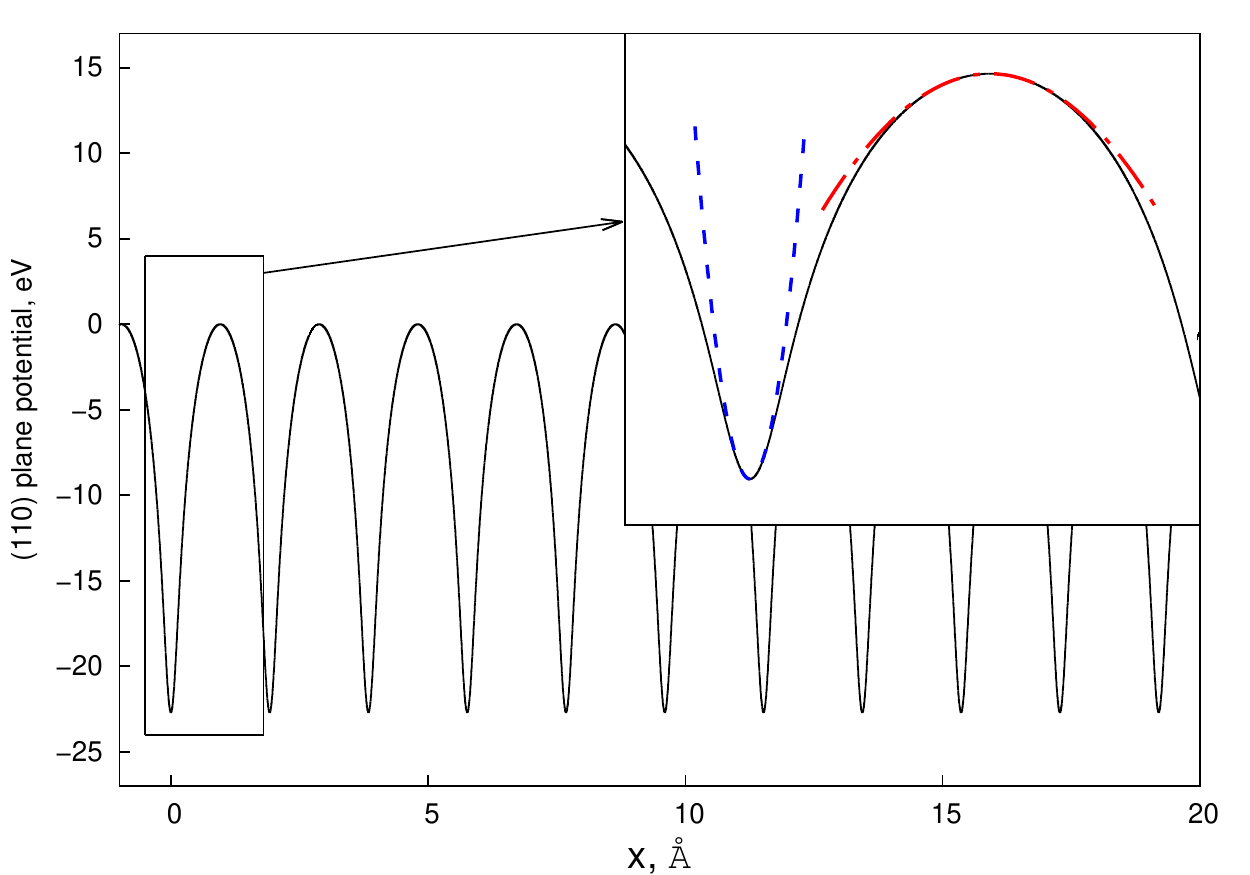}
\caption{Continuous potential of the plane (110) of Si crystal (solid line) and its approximations by quadratic functions of coordinates near extrema (blue dashed line -- approximation of the potential well for negative particles, red dash-and-dot line -- the one for positive particles, in this case is turned upside down)}
\label{fig_pot110}
\end{wrapfigure}

So, for both positively and negatively charged particles (PCP and NCP) in a crystal the continuous plane potential is a series of periodically placed potential wells and potential barriers. We must turn the potential upside down making the wells become barriers and vice versa if the sign of particle charge changes to the opposite (see Figure \ref{fig_pot110}).

In this case the potential in the neighbourhood of the bottom of the well for both PCP and NCP can be approximated by a parabola 
\begin{equation}\label{pot_fit}
U(x)=b (x - x_0)^2 + d,
\end{equation}
where the parameters $x_0$, $b$, $d$ may be found using a fitting procedure. The solution of equation of motion \eqref{eq_clasrho}
in the case of particles moving in such parabolic potential (in the case of parallel incidence) is a set of harmonic curves $x\!=\!x_0\,\cos{\omega t}$, where $\omega\!=\!\sqrt{2c^2 \left|b\right|/E_\parallel}$, therefore the spatial period of oscillations is 
\begin{equation}\label{T_osc}
T=\pi \beta\sqrt{2E_\parallel/\left|b\right|}\,,
\end{equation}
where $\beta=v/c$. We can define $T_0$ as the basic oscillation period that corresponds to a particle entering in the crystal in immediate proximity to the well bottom $x_0$. As we move away from $x_0$, the form of real potential deviates from the parabolic one, therefore the oscillation periods deviate from $T_0$. We can compare such a behaviour of NCP and PCP by analysing the difference between the real continuous potential and its fit by parabolas.

As far as in some vicinity of the well bottom the real potential remains close enough to its quadratic approximation, some part of particles falling into a crystal are ``focused" after scattering if the crystal thickness is a small integer number of half-periods $T_{\onehf}=T_0/2$ of the particle oscillations within the approximated parabolic potential \eqref{pot_fit}. As the thickness increases, the focusing effect weakens because of increasing difference of real coordinates of the oscillation nodes for different impact parameters. Obviously, the strongest focusing is observed at the first half-oscillation, $L=T_{\onehf}$. We can see in Figure\,\ref{fig_pot110} that the bottom of the potential well for PCP is approximated by a parabola much better than for NCP. It means that PCP will be focused more strongly and that the focusing effect will persist for a larger number of periods than for NCP (moreover, as stated below, the oscillation period for PCP is much larger than for NCP, so the thickness where the focusing effect can be observed is substantially  larger for PCP than for NCP that is caused by these two factors). 

For positively $(^+)$ and negatively $(^-)$ charged particles the parameters $b^\pm$ in the fit \eqref{pot_fit} of the potential of the (110) plane of Si crystal are $\left|b^+\right|=17.01\,\texttt{eV/\AA$^2$}$ and $\left|b^-\right|=407.6\,\texttt{eV/\AA$^2$}$. Therefore, we have 
\begin{equation}\label{T_osc_numbers}
\begin{array}{l}
T_{\onehf}^+=0.5385 \beta\sqrt{E_\parallel[eV]}\texttt{\AA},
\\
T_{\onehf}^- = \;\;\; 0.11 \beta\sqrt{E_\parallel[eV]}\texttt{\AA}.
\end{array}
\end{equation}
So, for this crystal plane the period for positively charged particles is almost 5 times larger than for negatively charged particles, that is only explained by the geometry of potential.

\begin{figure}[ptb!]\centering
 \begin{subfigure}{0.49	\textwidth}
\includegraphics [width=1\textwidth,natwidth=1063,natheight=1063]
{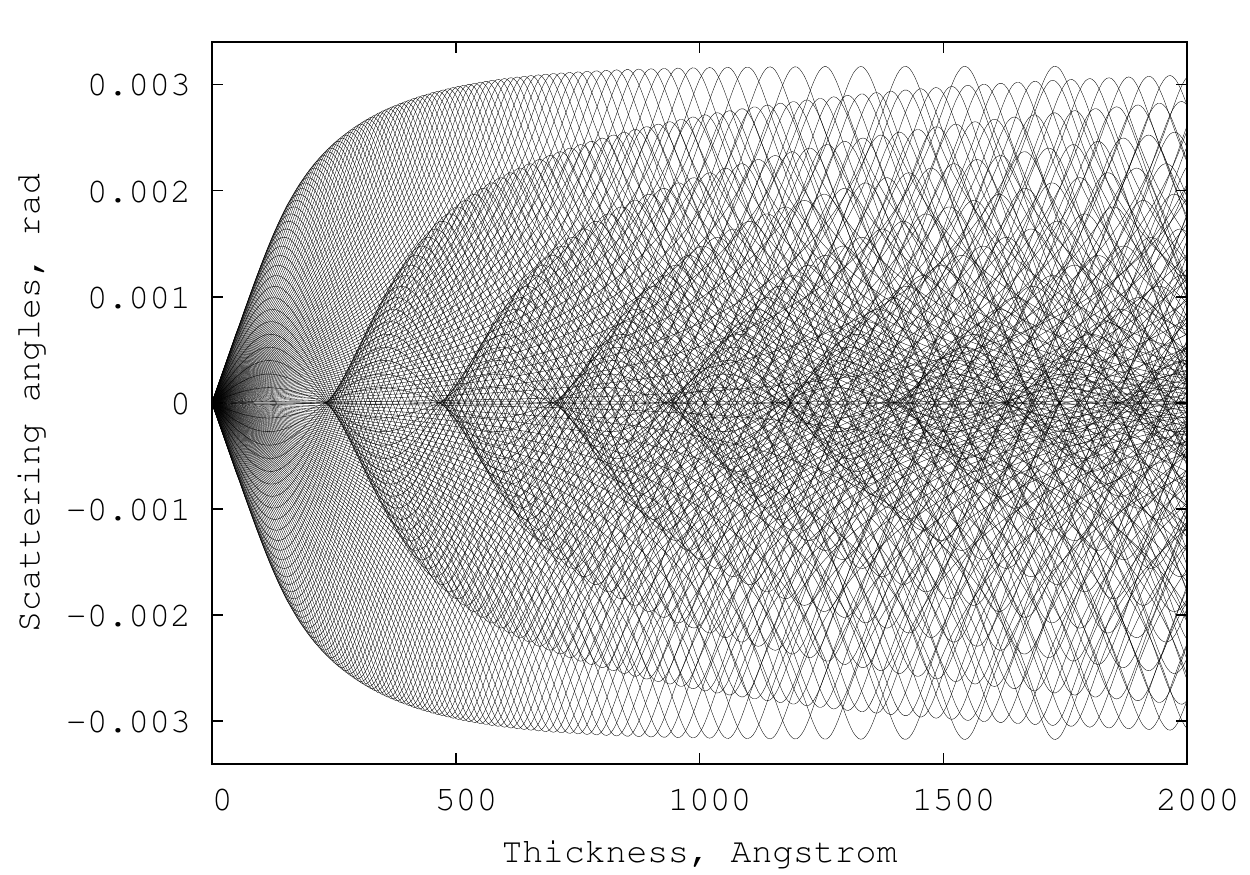}
\end{subfigure}
\hfill
\begin{subfigure}{0.49\textwidth}
\includegraphics [width=1\textwidth,natwidth=1063,natheight=1063]
{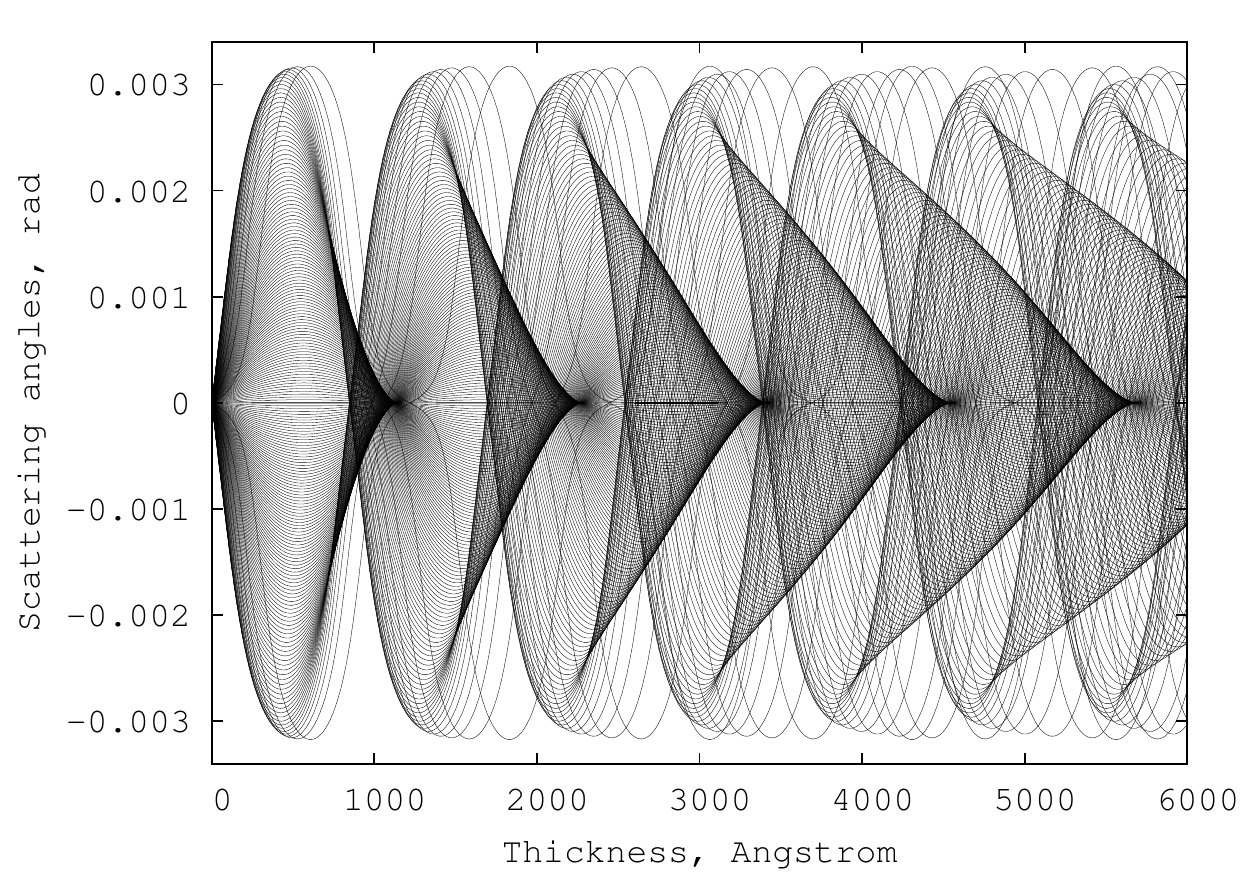}
\end{subfigure}
\caption{Classical scattering angles for different impact parameters of 4\,MeV electrons incident onto a 2000\,\AA~Si crystal parallel to (110) plane (left) and for different impact parameters of 4\,MeV positrons incident onto a 6000\,\AA~Si crystal parallel to (110) plane (right)}
\vspace{-2.ex}
\label{fig_caust_4MeV}
\end{figure}

From Figure\,\ref{fig_pot110} we see that, as far as we go away from the well bottom, the real potential curve for NCP becomes wider than its parabolic approximation and the one for PCP becomes narrower. This fact causes different symmetry of the scattering pictures for NCP and PCP in the neighbourhood of thicknesses $L=n\!\cdot\!T_{\onehf}$, where $n$ is an integer number, what is observed at comparison of both graphs of Figure \ref{fig_caust_4MeV}: we see that because of this the caustic lines for PCP cross among themselves, and the number of crossing caustics increases with thickness, that all not being observed for NCP.

\begin{figure}[pt]\centering
 \begin{subfigure}{0.46\textwidth}
\includegraphics [width=1\textwidth,natwidth=1063,natheight=1063]
{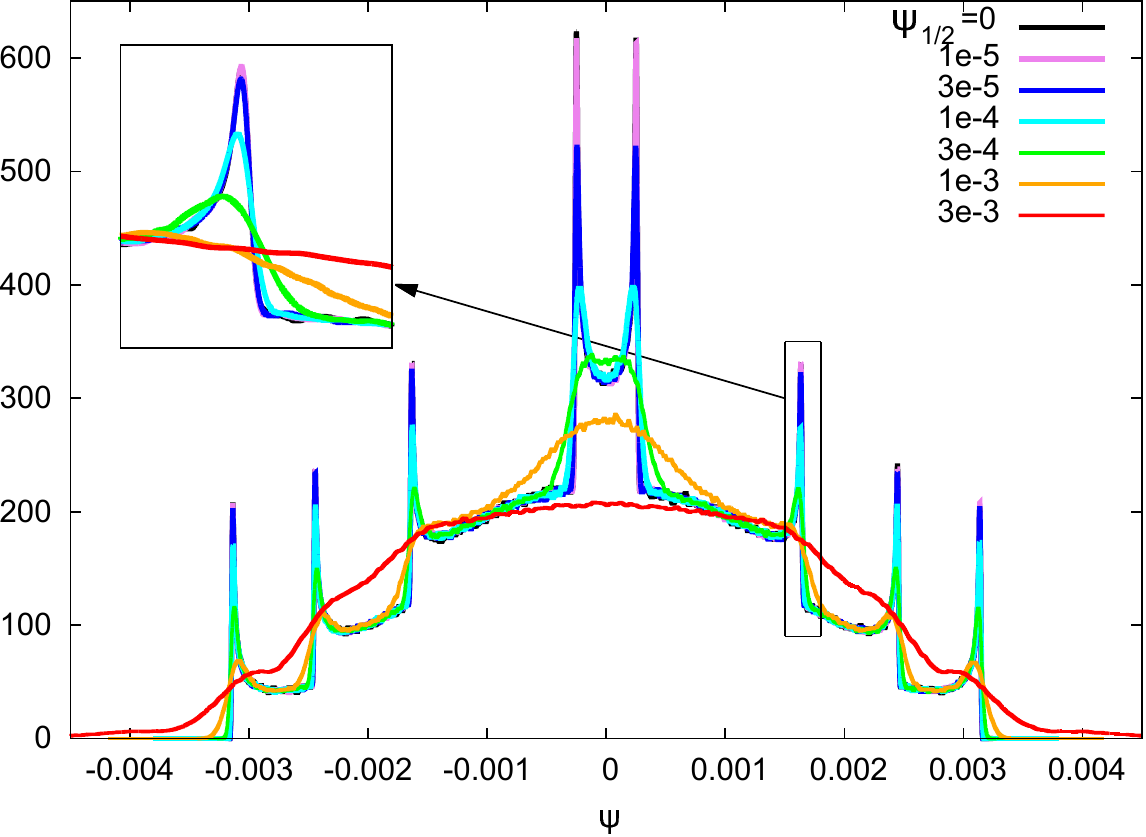}
\end{subfigure}
\hfill
 \begin{subfigure}{0.48\textwidth}
\includegraphics [width=1\textwidth,natwidth=1063,natheight=1063]
{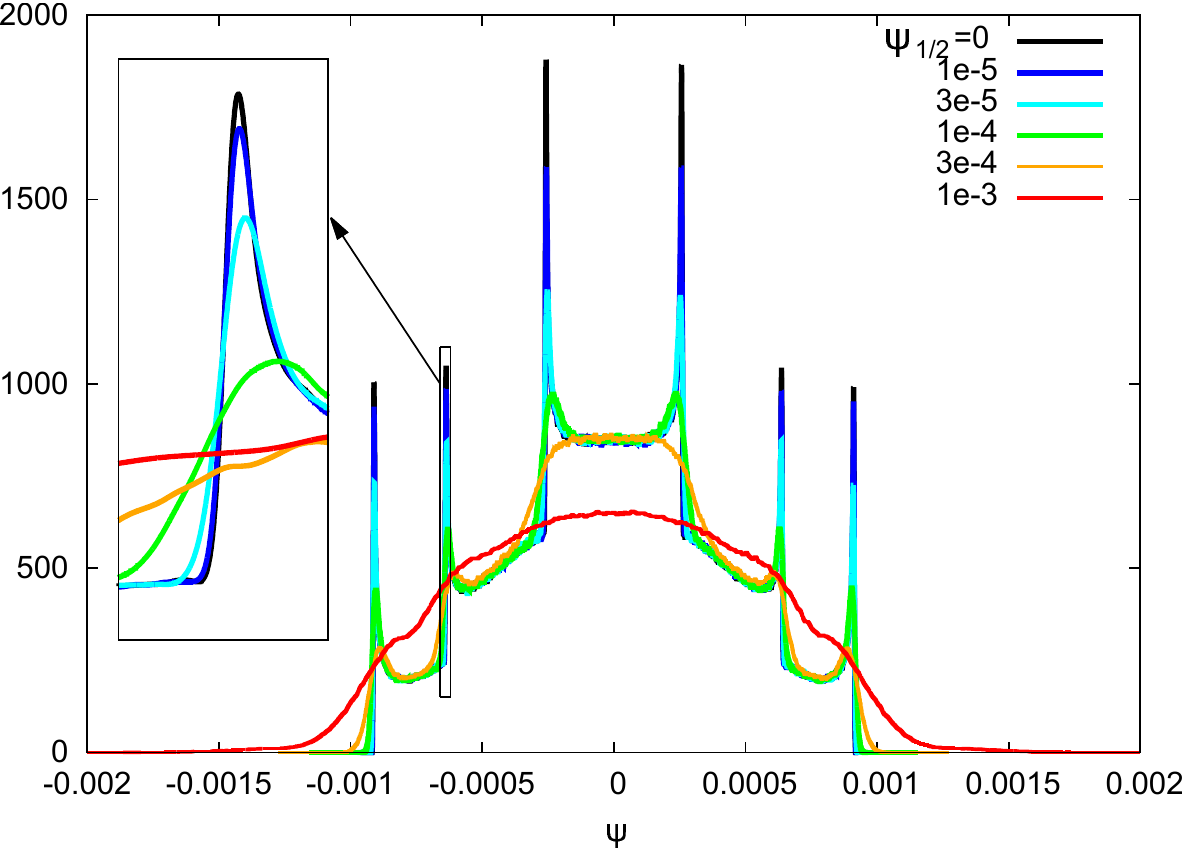}
\end{subfigure}
\caption{Classical simulation for scattering with different initial divergences for beams of 4\,MeV electrons by a 750\,\AA Si crystal (left) and 50\,MeV electrons by a 2000\,\AA~Si crystal (right), both along the (110) plane}
\vspace{-2.ex}
\label{fig_clas_ele_div}
\end{figure}

Figure \ref{fig_caust_4MeV} is a set of scattering angles of fast charged particles with a large number (200) of different impact parameters uniformly distributed throughout the full interval between planes. The maximal angular amplitude of oscillations in these graphs (along vertical axis) corresponds to the critical channelling angle,
\begin{equation}
\psi_c = \sqrt{2U_{\rm max}/E_\parallel}.
\label{eq_psicrit}
\end{equation}
The difference in these pictures is only caused by the asymmetry of planar potential relatively turnover upside down that is connected with the change of sign of the particle charge. We can see that entire scattering picture for PCP even changes its entire angular dimensions at first half-periods of oscillations, while in the case of NCP it quickly reaches its maximal value and then only changes its internal structure. Near each ``focusing point'' we see a caustic line enveloping the curves with similar impact parameters, coming out from this focusing point. As the thickness exceeds the ``focusing point'' the angular density having there a sharp maximum begins to bifurcate, and the two shown up maxima branch off in opposite directions, as observed in the plot of angular distributions of scattered electrons. The sections of the graphs of Figure \,\ref{fig_caust_4MeV} at constant thickness are proportional to the density of trajectories. By using these sections one can build the angular distributions of electrons scattered by crystal planes (see Figure\,\ref{fig_clas_ele_div}). Near the caustic lines, we have a strong increase of the density of lines from one side and abrupt decrease from the other. The angular distribution (Figure\,\ref{fig_clas_ele_div}) it is manifested as sharp maxima at corresponding angles, the angular distance between maxima being spread with the increase of thickness. We treat the presence of such sharp maxima as an appearance of the rainbow scattering phenomenon \cite{Nussenzv1992}. It, applied to axial channelling, is studied in works \cite{KrausNesk_PRB1986,Nesk_PRB1986,PetroNesk_PRA2013,NeskPero_PRL1987,MotapothNesk_PRB2012,Motapoth1}.

A real beam is not parallel, and, as far as we consider the angular properties of scattering, we must account for this fact. In order to do this, we performed a simulation of the scattering picture of charged particles of a beam with different initial angular spread (Figure\,\ref{fig_clas_ele_div}). The ideal situation of a parallel beam is also considered in order to see a ``pure'' picture. We can see that, as far as the initial beam spread increases, the rainbow lines disappear being smeared. We consider the angular distribution of the particles in the initial beam as gaussian, and characterize the angular spread using the parameter ``half-width-at-half-maximum" that means that at the angle corresponding to this parameter we have half of the maximal beam intensity (which is observed in center). We will designate this parameter as $\psi_{\onehf}$.

By performing a simulation and using such parameters for the description of a beam, we will see that in order to see the rainbow effect in classical study at scattering of, e.g., 4\,MeV electrons in 750\,\AA~Si crystal the beam angular spread must be at least not larger than $\psi_{\onehf} \sim 3{\cdot}10^{-4}-1{\cdot}10^{-3}$\,rad. The analogous condition for 50\,MeV electrons in the 2000\,\AA~Si crystal gives us $\psi_{\onehf}\sim 1{\cdot}10^{-4}-3{\cdot}10^{-4}$\,rad, being in both cases about $0.1-0.3\;\psi_c$.

So, as we can see, the conditions imposed on the beam quality in order to see the sharp maxima in classical study are strong enough, although we find them to be reachable.

\vspace{-2ex}
\section{Quantum scattering}
\vspace{-2ex}

In order to realize the quantum study we need to describe the initial beam as a wave packet, instead of the set of point-like particles as in classical case, and to study its development with time. As in the classical case, we must take into account that the initial beam has some spatial dimensions and angular spread. The analysis of the wave function will give us the information about the wave packet motion.

Within the quantum theory, a fast particle moving in a certain direction can be presented as a plane wave. A beam as a set of particles is therefore a set of plane waves, their directions of motion being distributed according to the laws of distribution of particles in the beam. We, however, find useful to describe mathematically the single particle wave function as a Gaussian wave packet
\begin{equation}
\Psi \left( x,t=0 \right) =\frac{1}{\sqrt {\sigma \sqrt{\pi\,}\,}}
\exp \left(- \frac{x^2}{2\sigma^2}+ i\frac{{p_x x}}{\hbar }\right),
\end{equation}
where the parameter $\sigma$ corresponds to the wave packet that covers a large number of neighbouring planes, hence, through the uncertainty relations, follows a very low angular divergence of such a wave packet.

In order to find the evolution of quantum state of the system with time we used the action of the time evolution operator onto the wave function. Such a way, with purpose to study the bound states levels, has been used in the works \cite{FeitFleck,Dabag1988,Dabag1988T1,Dabag1988T2,KozShuCherk2010,ShuSyshchNer2013}. 

The essence of our way of finding the evolution of wave packet is following: the change of the wave function with a time step $\delta t$ is obtained as a result of action of the time evolution operator onto its last step value:
\begin{equation}
\Psi \left( {x,t + \delta t} \right) = \exp \left( { - \frac{i}{\hbar }\delta t\,\hat H} \right)\Psi \left( {x,t} \right).
\label{psi_dt}
\end{equation}
But, we must take into account that the Hamiltonian of transverse motion is a sum of two \textit{non-commutating} terms
\begin{equation}
\hat H = -\frac{{{\hbar^2 c^2}}} {2{E_\parallel}} \frac{d^2}{dx^2}  + eU\left( {x} \right), 
\label{hamilt}
\end{equation}
that calls forth that we cannot present the exponent \eqref{psi_dt} of the hamiltonian \eqref{hamilt} as a simple consequent product of exponents. This does not let us take $\delta t$ as large as desired that would be in the case of absence of the potential, and we need to look for some approximation in order to get valuable results.  In order to perform the expansion of the exponent in \eqref{psi_dt} in series in terms of $\delta t$ we may use the Zassenhaus product formula. So, with precision up to terms proportional to $(\delta t)^3$, we have:
\begin{equation}
\begin{array}{l}
\exp \left( { - \frac{i}{\hbar}\,\delta t\hat H} \right) \approx
\\
\exp \left(-\frac{i}{2\hbar}eU\left(x \right)\delta t\right)
\exp \left(i\frac{\hbar c^2\delta t}{2E_\parallel} \frac{d^2}{dx^2}\right)
\exp {\left(-\frac{i}{2\hbar}eU\left(x\right)\delta t\right)}.
\end{array}
\label{}
\end{equation}
 In order to deal with the differential operator as an exponent index and not to calculate higher order derivatives we may take use of Fourier series formalism in which taking the derivative is reduced to the multiplication of each Fourier series term by a number. This procedure is exposed in the works \cite{FeitFleck, KozShuCherk2010, ShuSyshchNer2013}, and so on.
%

Once we have the wave function in position space we can take a Fourier transform in order to get it in momentum space. Therefore, by taking the square of its absolute value we obtain the angular distribution of scattered particles in quantum case. So, the probability for the particle scattering in the interval $[\psi,\psi+d\psi]$ is
\begin{equation}
dw(\psi)=\left|\int{\Psi\left(x,t\!=\!L/v\right)e^{-ip_{\parallel}\psi x/\hbar}dx}\right|^2 \frac{p_\parallel d\psi}{2\pi\hbar}.
\label{}
\end{equation}

\begin{figure}[pt]\centering
 \begin{subfigure}{0.49\textwidth}
\includegraphics [width=1\textwidth,natwidth=1063,natheight=1063]
{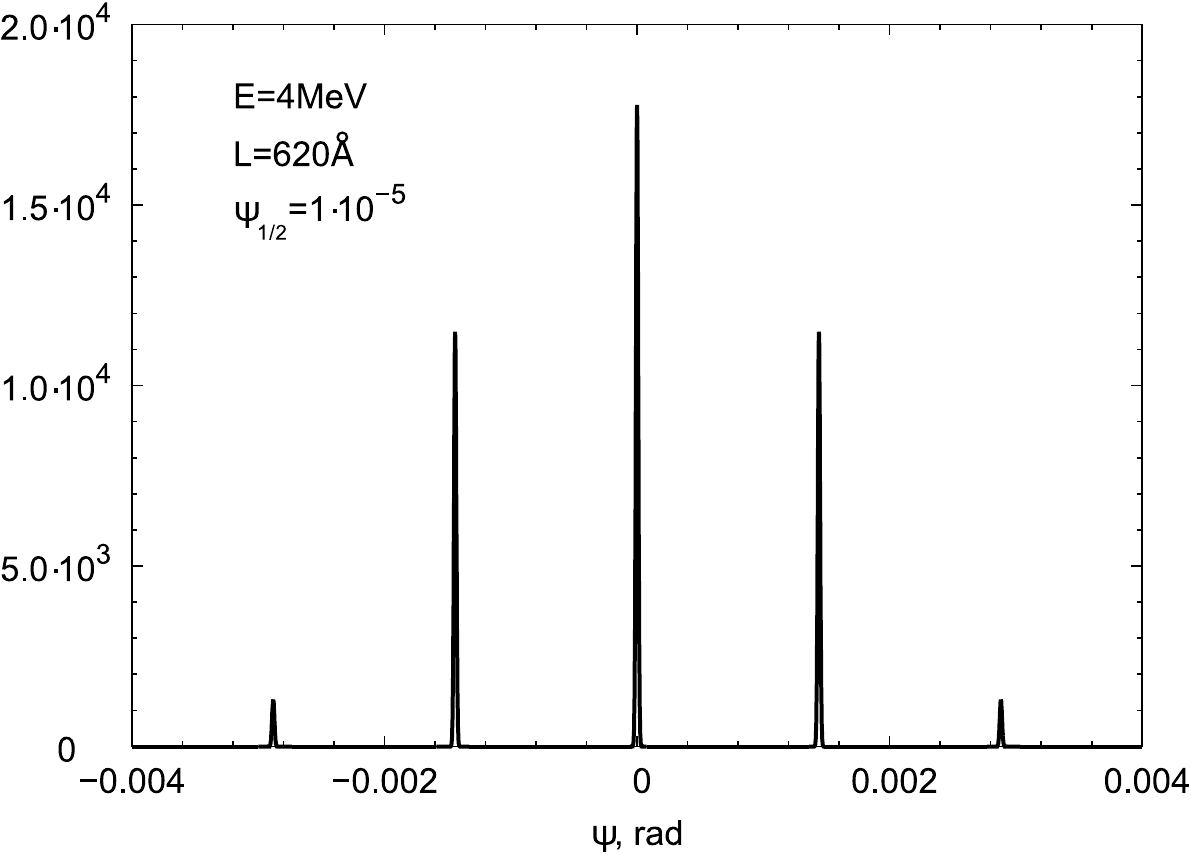}
\end{subfigure}
 \begin{subfigure}{0.49\textwidth}
\includegraphics [width=1\textwidth,natwidth=1063,natheight=1063]
{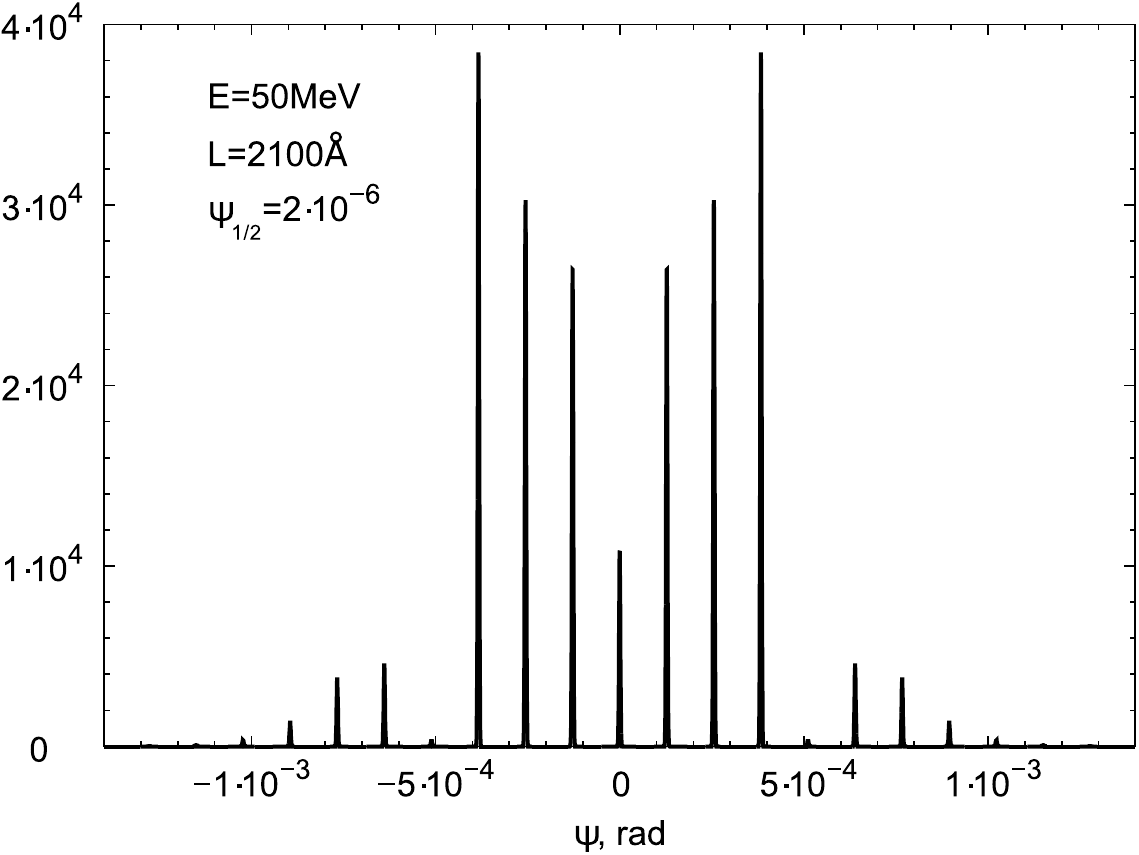}
\end{subfigure}
\caption{Diffraction of single electrons on (110) planes of Si crystal}
\vspace{-2.3ex}
\label{fig_quant_SolEle}
\end{figure}

In our study the wave functions describing single particles correspond to almost plane waves with the divergence of about $\psi_{\onehf}\sim0.001\psi_c$. In Figure\,\ref{fig_quant_SolEle} we can see that the diffraction picture of single electrons in crystal is a row of $\delta$-like maxima. The set of equidistant narrow maxima can be explained as the expansion in reciprocal lattice vectors. We can represent the particles as waves with the wavelength equal to the de Broglie length $\lambdabar=\hbar/p_\parallel$, therefore the angles corresponding to the maxima must satisfy the relation 
\begin{equation}\label{eq_reciprlat}
\psi_n=g_n/p_\parallel,
\end{equation}
where $g_n=2\pi\hbar n/d_x$ is the reciprocal lattice vector, $n$ -- integer number and $d_x$ is the distance between the crystl planes. By the other words, we consider the crystal as a diffraction grating and the particles scattering -- as a scattering of de Broglie waves on this grating, so the sharp maxima present on Figure\,\ref{fig_quant_SolEle} we explain as the manifestation of the interference of electrons with themselves at scattering on different planes. Higher is the particle energy, more densely the allowed angles for particle scattering are situated. As far as the distance between the peaks is proportional to $1/E$ \eqref{eq_reciprlat} and the channelling angle $\psi_c\propto 1/\sqrt{E}$, for higher energies we have the number of quantum peaks inside the scattering range increasing proportional to the square root of ${E}$. We can estimate the number of peaks: $N_p\approx \psi_c/\psi_T=\frac{\sqrt{2U_{\rm max}}d}{2\pi c\hbar}\sqrt{E}$. For the $(110)$ plane of Si crystal it is approximately $N_p\sim\sqrt{E[MeV]}$.

We get the results for scattering in quantum case by averaging contributions of solitary wave packets of single charged particles by summing up their contributions with Gauss distribution function that modulates the beam divergence:
\begin{equation}\label{eq_beamavrg}
\overline{w_{beam}(\psi)}=\frac{1}{\sigma_b\sqrt{\pi}} \int{e^{-\psi_i^2/\sigma_b^2}}\,w(\psi_i,\psi)\,d\psi_i,
\end{equation} 
where $\overline{w_{beam}(\psi)}$ states for the density of probability of scattering of the beam incident as a whole parallel to the crystal planes in the direction $\psi$, and $w(\psi_i,\psi)$ is the density of probability that the particle incident at the angle $\psi_i$ to the planes is scattered in the direction $\psi$.
\begin{figure}[t]
\centering
 \includegraphics [width=1\textwidth,natwidth=1063,natheight=1063]
{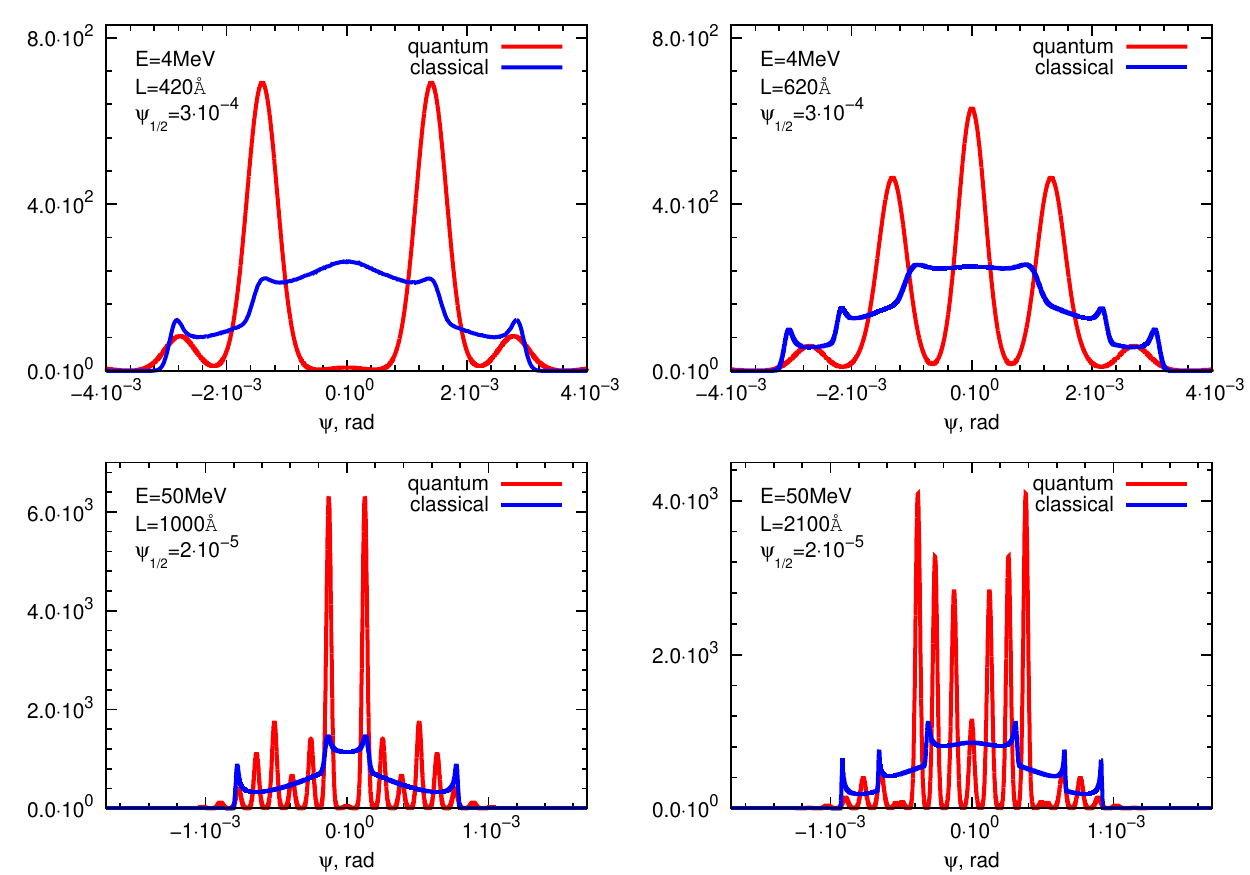}
\caption{Quantum and classical simulations of angular distribution of 4\,MeV (above) and 50\,MeV (below) electrons  scattered in a Si crystal incident parallel to the ${(}110{)}$ plane}
\label{fig_QuCl_different}
\vspace{-2.3ex}
\end{figure}

It is possible to compare quantum results for angular distributions with the classical ones at the same parameters. In Figure~\ref{fig_QuCl_different} we present quantum angular distributions obtained by the method described here and the classical ones as solutions of the classical equation of motion \eqref{eq_clasrho}. We see that, lower is the particle energy, greater is the difference between the classical and quantum pictures. Particularly for low energies, the classical and quantum pictures may be substantially different. Besides, the averaging over a diverging beam in quantum case can make neighbouring maxima flow together, make them disappear or displace them. For example, on the picture for 420\,\AA~we see that the central part of the angular distribution of scattering, being elevated in classical distribution, 	is strongly depressed in quantum case. From the other hand, for higher energies, the sharp maxima observed in classical pictures have some manifestation in quantum case: for the angles that correspond to them the quantum peaks are higher, and for GeV energies they flow together so as quantum picture approaches the classical one. 

As written above, the positions of quantum and classical maxima have different nature, so the positions of quantum ones do not depend on the crystal thickness (at the absence of strong influence of the beam divergence), whereas the classical ones migrate towards outside of the scattering figure with the increase of thickness, so any coincidence of classical and quantum maxima is accidental. It can be observed at one crystal thickness and not be observed for another one.

\vspace{-2ex}
\section{Conclusion}
\vspace{-2ex}

In this paper we propose an idea for experiment on planar scattering of fast charged particles in ultrathin crystal, in order to observe quantum and classical effects that can manifest themselves in scattering picture. If the quality of beam and crystal is good enough and the resolution of detectors is high enough we expect the observing of a series of spots of different brightness that, by varying the initial parameters may be referred to manifestation of quantum or classical nature of processes at interaction of charged particles with ultrathin crystal. 

The obtained results for angular distributions of the electrons scattered by a ultrathin crystal show that, for Si crystal with thickness about several hundreds of \AA~the quantum effects at scattering can be essential, that are connected with the interference effect of single electrons on a set of crystal planes. This effect is particilarly bright for the electrons energies about a few MeV. As the plots on Figure~5 show, the electron beam must have parameters attainable on the PhIl and ThomX facilities in the laboratory LAL in Orsay, France.

We did not include in this paper the study of the levels of transversal energy of particles that they occupy in the potential wells of crystal potential. These energy levels are to be observed by using other technical measures, such that will let us register the photons irradiated at interaction of charged particles with crystal, so such observations could reply the question about the mechanisms of arising of these levels and their nature, whether they are connected only with the potential well form or also with the reciprocal influence of neighbouring crystal planes and interference of charged particles on them. These questions could be answered by performing, in addition and in connection with the study proposed in this paper, of another study of radiation arising at interaction of charged particles with ultrathin crystal.

\vspace{-2ex}
\section{Acknowledgements}
\vspace{-2ex}

Research conducted in the scope of the IDEATE International Associated Laboratory (LIA). The work was done with partial support of the NAS of Ukraine, project $\Phi$-12 and of Ministry of Education and Science of Ukraine (project no.~0115U000473). One of the authors (S.N.\,Shul'ga) is grateful to the Laboratoire de l'Acc\'el\'erateur Lin\'eaire where the essential part of this work was done and its researchers for the hospitality and fruitful discussions.


\end{document}